\documentclass{ws-mpla}

\usepackage[super]{cite}
\usepackage{xcolor}
\usepackage[verbose,hypertexnames=false]{hyperref}
\hypersetup{colorlinks=false,allbordercolors=blue,pdfborderstyle={/S/U/W 1}}

\begin{document}
	
	\markboth{Ghosh, Paul, Sharma and Chanda}{Physical properties and the maximum compactness bound of a class of compact stars in $f(Q)$ gravity}
	
	\catchline{}{}{}{}{}
	
	\title{Physical properties and the maximum compactness bound of a class of compact stars in $f(Q)$ gravity}

	\author{Arpita Ghosh\footnote{E-mail: arpitaghosh92727@gmail.com}, Abhishek Paul\footnote{E-mail: paulabhishek.res@gmail.com}, Ranjan Sharma\footnote{E-mail: rsharma@associates.iucaa.in (Corresponding author)}~ and Samstuti Chanda\footnote{E-mail: schanda93.dta@gmail.com}}
	
	\address{IUCAA Centre for Astronomy Research and Development ICARD),\\ Department of Physics, Cooch Behar Panchanan Barma University,\\
		Vivekananda Street, Cooch Behar 736101, India.}

	\maketitle
	
	\begin{history}
		\received{(Day Month Year)}
		\revised{(Day Month Year)}
		\accepted{(Day Month Year)}
		\published{(Day Month Year)}
	\end{history}
	
\begin{abstract}

	{\bf Motivation:} Motivated by the growing interest in understanding the role of non-metricity in describing dense stellar systems, in this paper, we study compact stellar configurations	within the framework of linear $f(Q)$ gravity.\\
	
\noindent {\bf Methodology:} By adopting a linear modification of the form $f(Q) = \alpha Q+\beta$, we analyze the internal structure and physical properties of an anisotropic relativistic star within the framework of $f(Q)$ gravity. We employ the Karmarkar's condition together with the Vaidya-Tikekar metric ansatz to obtain a closed-form interior solution of the star. The interior solution is then matched to the Schwarzschild exterior solution across the boundary of the star. By varying the model parameters, we analyze physical features of the resultant stellar configuration.\\
	
\noindent {\bf Results:} We note distinctive features in the density, pressure, anisotropy and total mass of the star under a such modification. By enforcing the condition that the central pressure remains finite, we obtain the maximum compactness bound which is shown to depend solely on the Vaidya-Tikekar curvature parameter $K$. We recover the Buchdahl bound for the curvature parameter $K=0$, which corresponds to the solution for an isotropic and homogeneous fluid sphere. Utilizing the energy density and radial pressure profiles, we numerically integrate the modified Tolman-Oppenheimer-Volkoff equations and obtain the mass-radius ($M-R$) relationships for different values of the model parameter $\alpha$. We note that for higher values of $\alpha$, the maximum mass and radius decrease, shifting the stable branch towards ultra-compact configurations. An interesting observation in our analysis is that a linearly modified $f(Q)$ gravity model can support comparatively low mass stars. Utilizing the observed mass of the pulsar $XTE~J1814-338$, we demonstrate how our model can be used to fine-tune the radius of the star.
\end{abstract}
	
	
	\section{Introduction}
	\label{sec1}
	
	The General Theory of Relativity (GTR) developed by Einstein remains the most modern theory of gravity till date as the theory has successfully predicted many tests of gravity including the most recent discovery of gravitational waves. Despite its remarkable success, GTR faces many challenges on small and large scales. Among many other issues, one of the biggest challenges that GTR faces is its limitation in explaining the current accelerated expansion of the universe vis-a-vis the hidden source of the extra repulsive pressure. To explore the possible sources of late time cosmic expansion, different modified theories of gravity have so far been developed \cite{Ferraro2008,Geng2011,Cai2011,Jarv2016}. Such modifications are mostly done in the matter segment of the Einstein-Hilbert action as well as in the geometry of the associated spacetime. Among many alternative theories, one of the most popular proposals is the $f(R)$ theory of gravity in which the Ricci scalar ($R$) is replaced by an arbitrary function $f(R)$ in the Einstein-Hilbert (E-H) action \cite{Nojiri2007}. Note that GTR is based on the Riemannian geometry, where one considers the Ricci curvature $R$ as a fundamental property of spacetime. However, this prescription does not include torsion or non-metricity, in general. A non-Riemannian geometry can incorporate torsion and non-metricity as additional geometrical properties of the spacetime. Teleparallel gravity is an alternative to GTR, where the gravitational forces arise from the torsion $T$ instead of the curvature $R$. As an extension of the symmetric teleparallel gravity, a modified $f(Q)$ gravity can be developed \cite{Harko2018}, where the non-metricity serves as a mediator to gravitational interactions \cite{Jimenez201801}. It turns out that the choice of non-zero metricity can be an alternative approach to explain the late cosmic expansion \cite{Kirsch2005}. Consequently, of late, $f(Q)$ gravity has gained significant attention in understanding its cosmological and astrophysical implications. Hohmann {\em et al} \cite{Hohmann2019} examined the propagation velocity and potential polarization of gravitational waves in $f(Q)$ gravity. Soudi {\em et al} \cite{Soudi2019} studied the strong field behaviour by analyzing the gravitational wave polarization. Various investigations have been carried out to explore the implications of  $f(Q)$ gravity in constraining observational data \cite{Lazkoz2019,Ayuso2021} and its effects on energy conditions \cite{Mandal2020}, cosmography \cite{Mandal20201}, bouncing scenarios \cite{Mandal2021,Bajardi2020} and black hole physics \cite{Ambrosio2022}. For a comprehensive review of $f(Q)$ gravity theory, we refer to  \refcite{Heisenberg2023,Nguyen2021,Esposito2022,Zhao2022,Mustafa2021,Kimura2021,Dimakis2021} and references therein.  
	
	Even though $f(Q)$ gravity is primarily motivated by cosmological observations, the theory has gained widespread attention in astrophysics in the recent past. In particular, such a modification provides ample motivation to study its effects on relativistic, highly compact stellar models. Assuming the Buchdahl metric potential \cite{Buchdahl1959} for a strange star, Sokoliuk {\em et al} \cite{Sokoliuk2022} analyzed the physical features of a compact star in both linear and non-linear forms of modified symmetric teleparallel gravity. Under a linear $f(Q)$ regime, Maurya and collaborators \cite{Maurya2022,maurya1fq} explored several aspects of a compact star structure. In particular, they have analyzed the maximum permissible mass of an anisotropic strange quark star and showed that the theory permits heavier configurations consistent with the GW190814 observation \cite{Maurya2022}. They further analyzed anisotropic stellar models in $f(Q)$ gravity capable of reaching $2.08$–$2.67\,M_{\odot}$, potentially entering into the GW190814 predicted mass gap and observed that the associated radii should be larger than those obtained in GTR~\cite{maurya1fq}. They also investigated how the $f(Q)$ gravity parameter could influence on Korkina-Orlyanskii and Buchdahl models, demonstrating its impact on stability and mass-radius trends~\cite{mauryafq2} and presented an exact anisotropic compact star solution satisfying all physical criteria. The model was shown to be consistent with the $Her~X-1$ data, where certain parameter combinations was shown to lead to a marked enhancement in the mass distribution \cite{mauryafq3}. Errehymy \textit{et al}~\cite{errfq1} developed anisotropic compact star solutions in linear $f(Q)$ gravity and demonstrated that suitable parameter choices mimic polytropic or quark matter behavior, producing high mass stars consistent with GW190814/GW200210, with radii of $\sim 11-12~km$ and large moments of inertia. On the other hand, Bhar \textit{et al}~\cite{bharfq1} extended the linear $f(Q)$ gravity to dark energy supported compact stars. Their model yields larger and heavier stellar configurations with a predicted maximum mass of $2.57\,M_\odot$ consistent with the GW190814 event observation. Kumar \textit{et al}~\cite{kumarfq1} obtained a charged, isotropic compact star solution by incorporating the Buchdahl metric ansatz in $f(Q)$ gravity and showed that it could provide a viable description of realistic stars such as $Her~X-1$. Lin and Zhai \cite{Lin2022} analyzed the implications of $f(Q)$ gravity in the case of a spherically symmetric stellar configuration. In their paper, with $f(Q)=Q+\alpha Q^{2}$, for a polytropic star, they have shown that a negative modification $(\alpha < 0)$ could provide more stellar masses. In contrast, a positive value could reduce the amount of matter within the star. Making use of the Tolamn-Kuchowicz ansatz \cite{Tolman1968} for a hybrid star composed of strange quark matter together with baryonic matter, Bhar {\em et al} \cite{Bhar2023} analyzed the maximum mass-radius relationship in $f(Q)$ gravity. Araujo and Fortes  \cite{Araujo2407} investigated the maximum mass limit of a compact star admitting a polytropic EOS in $f(Q)= Q + \xi Q^{2}$ gravity. Alwan {\em et al} \cite{Alwan2024} have studied neutron stars in covariant $f(Q)$ gravity. Making use of Karmarkar's condition together with an ansatz for the metric potential $g_{rr}$, Paul {\em et al} \cite{SatPaul2025} developed an anisotropic stellar model in $f(Q)$ gravity. A number of authors \cite{Mandal20222,Errehymy2022,Gul202401} have also studied the dynamical behaviour of compact stars.
	
	It should be stressed here that physical features of compact stars have also been probed in other modified theories of gravity which include the works in $f(R)$ gravity \cite{Kobayashi2008,Upadhye2009}, $f(R,T)$ gravity \cite{Nashed2023,Asghar2023,Naseer2024,Murshid2024,Bhattacharyaa2024, Hansraj_24}, $f(T)$ gravity \cite{Aftergood2014,DeBenedictis2016}, $f(G)$ gravity \cite{Abbas2015,Malik2022}, $f(G,T)$ gravity \cite{Shamir2017,Sadiq2022}, Rastall gravity \cite{Bhar2020}, $f(Q,T)$ gravity \cite{Das2024} and Teleparallel Palatini theory \cite{Jimenez2018}. Using the minimal geometric deformation technique, Pradhan {\em et al} \cite{Pradhan2024} studied geometrically deformed compact objects in $f(Q,T)$ gravity in the presence of an electric field. Gul {\em et al} \cite{Gul2024} explored the viability and stability of compact stellar objects characterized by anisotropic matter within the framework of $f(Q,T)$ gravity. 
	
The main objective of the current investigation is to develop a compact stellar model in $f(Q)$ gravity the physical behavior of which can be easily tractable. By assume a linear for of the function $f(Q) =\alpha Q +\beta$, we utilize the Karmarkar's condition together with the Vaidya-Tikekar metric ansatz, in this paper, we analyze the implications of pure $f(Q)$ gravity on the gross physical behaviour of an anisotropic compact stellar configuration, with particular focus on how non-metricity influences the physical properties, the maximum compactness bound and mass-radius relationship. Our analysis will, hopefully, enrich the current understanding of the impacts of modified $f(Q)$ gravity theories in the linear regime constructed within an embedding class-I framework. Recent theoretical studies indicate that the linear choice $f(Q)=\alpha Q+\beta$ occupies a privileged and physically consistent position within symmetric teleparallel gravity. In particular, De and Loo ~\cite{De2023} showed that only linear $f(Q)$ models, in general, ensure covariant conservation of the energy-momentum tensor in arbitrary spacetime geometries. However, a non-linear $f(Q)$ modification violates this condition unless the non-metricity scalar $Q$ is assumed to be a constant, in which case the theory dynamically yields  general relativistic results with an effective cosmological constant. Heisenberg and Pastor-Marcos~\cite{Heisenberg2025} demonstrated that for compact objects such as neutron stars, non-linear extensions including $f(Q)=Q+\alpha Q^{2}$ and $f(Q)=Q^{\beta}$ tend to recover GR-like dynamics under standard regularity and asymptotic assumptions, unless the affine connection is treated as an active dynamical degree of freedom. These results suggest that physically meaningful deviations from GTR in non-linear $f(Q)$ gravity require non-trivial connection dynamics or departures from standard boundary conditions. In the context of astrophysics, this requires a more careful analysis of modified $f(Q)$ gravity models and demands further probe. Meanwhile, by adopting a linear $f(Q)$ theory as a self-consistent baseline, in this paper, we intend to analyze its implications in stellar modelling.

	In our work, we first develop an interior solution describing an anisotropic stellar configuration in $f(Q)$ gravity whose exterior spacetime is described by the vacuum Schwarzschild metric. Note that a relaxation in the pressure isotropy condition is found to be relevant in the studies of compact stars as pointed out by Herrea \cite{Herrera2020}. Various factors such as presence of magnetic field, phase transition, viscosity, rotation and the mixture of two fluids, among others, can be potential sources for such an anisotropy \cite{Ruderman1972,Bower1974,Bondi1992,Herrera1997}. In fact, even an isotropic star may develop anisotropy during its evolution due to internal physical processes, as suggested by Herrera  \cite{Herrera2020}.  Amongst many others, anisotropic stellar distribution has been investigated by Sharma and Maharaj \cite{Sharma2007} and Maurya {\em et al} \cite{Maurya2015,Maurya2018} where effects of anisotropy on the physical properties of compact stars have been dealt with in details.
	
	Our paper is organized as follows: Section $2$ provides the mathematical framework of $f(Q)$ gravity. Corresponding to a static and spherically symmetric anisotropic stellar configuration, we lay down the subsequent field equations in this section. In Section~$3$, making use of the Karmarkar's condition \cite{Karmakar1948,Abbas2019,Sharif2022} together with a particular ansatz for one of the metric potentials, we develop an anisotropic stellar model in $f(Q)$ gravity. Using the relevant matching conditions in Section~$4$, we fix the values of the model parameters. Gross physical properties of the resultant stellar configuration are studied in Section~$5$. In Section~$6$, we determine the maximum compactness bound of the resultant configuration, which might be treated as $f(Q)$ analogue of the Buchdal bound in GTR. The mass-radius relationship of the configuration is studied and its relevance in the context of some recent observations is analyzed. Some concluding remarks are made in Section~$7$.
	\section{$f(Q)$ gravity: General formalism}
	\label{sec2}
	\subsection{General formalism}
	One formulates the $f(Q)$ gravity model by setting the curvature $R$ and the torsion $T$ to be zero in the non-Riemannian geometry, assuming that the geometric information is encoded in the non-metricity $Q$ only \cite{Mustafa2023}. For the action 
	\begin{equation}
		S=\int \sqrt{-g} \ d^{4}x \left[\frac{1}{2} f(Q)+ \lambda^{bm n}_{a} R^{a}_{bmn}+ \lambda_{a}^{mn} T^{a}_{mn}+L_{m}\right],
		\label{eq1}
	\end{equation}
	the non-metricity $Q$ is defined in terms of the affine connections as
	\begin{equation}
		Q_{au n } \equiv \nabla_{a}g_{u n} = \partial_{a} g_{u n}- \Gamma^{\varrho}_{au} g_{\varrho n} - \Gamma^{\varrho}_{an} g_{u\varrho},
		\label{eq2}
	\end{equation}
	where the affine connections take the form
	\begin{eqnarray}
		\Gamma^{\varrho}_{u n} &=& \left\{ ^{\varrho}_{u n}\right\}+K^{\varrho}_{u n} +L^{\varrho}_{u n},\\
		\left\{ ^{\varrho}_{u n}\right\}&=&\frac{1}{2}g^{\varrho b}\left( \partial_{u}\ g_{bn}+\partial_{n}\ g_{bu}-\partial_{b}\ g_{u n}\right),\\
		K^{\varrho}_{u n}&=&\frac{1}{2}T^{\varrho}_{u n} + T_{\left(u\right.\left.\  n \right)}^{ \ \ \varrho},\\
		T^{\varrho}_{u n} &\equiv& 2\Gamma^{\varrho}_{\left[u n\right]},\\
		L^{\varrho}_{u n} &=& \frac{1}{2}Q^{\varrho}_{u n} - Q_{\left(u\right.\left.\  n \right)}^{ \ \ \varrho},
		\label{eq3to5}
	\end{eqnarray}
expressed in terms of Levi-Civita connection $\left\{ ^{\varrho}_{u n}\right\}$, the contortion tensors $K^{\varrho}_{u n}$, disformation tensor $L^{\varrho}_{u n}$ and torsion tensor $	T^{\varrho}_{u n}$.	In equation (\ref{eq1}), $\lambda^{bmn}_{a}$ corresponds to multipliers of the Lagrangian and $L_{m}$ is the Lagrangian matter density. 
	Expressing the non-metricity conjugate as
	\begin{equation}
		P^{a}_{u n} = -\frac{1}{4} Q^{a}_{u n}+\frac{1}{2} Q_{\left(u\right.\left.\  n \right)}^{ \ \ a}+ \frac{1}{4} \left( Q^{a}-\bar{Q^{a}}\right)\ g_{u n}  - \frac{1}{4}\delta^{a} \  _{\left(u\right.} Q \ _{\left. n \right)},
	\end{equation}
	we have the non-metricity scalar
	\begin{equation}
		Q = - Q_{au n}P^{{au n}},
		\label{eq9}
	\end{equation}
	where, $\bar{Q}_{a} \equiv Q_{a u}^{  u}, \ \ Q_{a} \equiv   Q_{a \ u}^{\ u}$.
	
	The subsequent field equations are obtained by  varying the action ($\ref{eq1}$), which yields 
	\begin{equation}
		-T_{u n} = \frac{2}{\sqrt{-g}} \nabla_{a} \left(\sqrt{-g} \ f_{Q} \ P^{a}_{u n}\right) +   \frac{1}{2} \ g_{u n} \ f + f_{Q} \ \left( P_{u a b} \ Q^{ab}_{n} - 2 \ Q_{abu} \ P^{ab}_{n}\right),
		\label{eq10}
	\end{equation}
	where $f_{Q}= \partial_{Q}\ f(Q)$. Further, varying equation ($\ref{eq1}$) with respect to the affine connections, we obtain 
	\begin{eqnarray}
		\nabla_{\rho} \ \varrho_{a}^{nu\rho} + \varrho_{a}^{u n} &=& \sqrt{-g} \ f_Q p_{a}^{u n} + H^{u n}_{a},
		\label{eq12}
	\end{eqnarray}
	where the energy-momentum tensor has the form 
	\begin{equation}
		T_{u n} = - \frac{2}{\sqrt{-g}} \frac{\partial \left( \sqrt{-g} \ L_{matter}\right)}{\partial \ g^{un}}.
		\label{eq11}
	\end{equation}
	The density for the hyper-momentum tensor has the form 
	\begin{equation}
		H^{u n}_{a} = -\frac{1}{2} \frac{\delta\ L_{matter}}{\delta \Gamma^{a}_{u n}}.
		\label{eq13}
	\end{equation}
	Now, using the asymmetric property of $n$ and $u$, equation ($\ref{eq12}$) can be written as
	\begin{equation}
		\nabla_{u}\nabla_{n} \left( \sqrt{-g} \ f_{Q} \ P^{u n}_{a} + H^{un}_{a} \right) =0.
		\label{eq14}
	\end{equation}
	Using the relation $\nabla_{u}\nabla_{n} \ H^{u n}_{a} = 0$, we write equation ($\ref{eq14}$) as
	\begin{equation}
		\nabla_{u}\nabla_{n} \left( \sqrt{-g} \ f_{Q} \ P^{un}\right) = 0.
		\label{15}
	\end{equation}
	
	In a particular coordinate system, in the case of coincident gauge, the non-metricity is reduced to the following expression
	\begin{equation}
		Q_{au n} = \partial_{a} \ g_{u n},
		\label{eq17}
	\end{equation}
	where, in the absence of curvature or torsion, the affine connections are given by
	\begin{equation}
		\Gamma^{a}_{u n} = \left( \frac{\partial x^{a}}{\partial \xi^{\varrho}}\right) \partial_{u} \ \partial_{n} \xi^{\varrho}.
		\label{eq16}
	\end{equation}
	In the following sub-section, for a given line element, we obtain the non-metricity scalar and subsequent field equations.  
	
	\subsection{The line element}
	To understand the physical behaviour of a spherically symmetric, static and compact stellar object, we assume that the following line element describe the internal geometry of the star
	\begin{eqnarray}
		ds^{2} &=& -e^{\nu(r)} \ dt^{2}+e^{\lambda(r)} \ dr^{2}+ r^{2} \left( d\theta^{2} + \ \sin^{2} \theta \ d\phi^{2}\right),
		\label{eq18}
	\end{eqnarray}
	where $\nu(r)$ and $\lambda(r)$ are the undetermined functions of the radial coordinate $r$. Substituting equation ($\ref{eq18}$) into ($\ref{eq9}$), we obtain the non-metricity scalar $Q$ as
	\begin{equation}
		Q = - \frac{\left(2 \ e^{-\lambda(r)}\right)\ \left(\nu'(r)+\frac{1}{r}\right)}{r},
		\label{eq19}
	\end{equation}
	where a prime denotes a derivative with respect to $r$. The energy-momentum tensor for an anisotropic matter is assumed to be of the form
	\begin{eqnarray}
		T_{ij} &=& \left( \rho + p_{t}\right)\zeta_{i}\zeta_{j} - p_{t} \ g_{ij} + \left( p_{r}-p_{t}\right)\xi_{i}\xi_{j},
		\label{eq20}
	\end{eqnarray}
	where $\rho, p_{r}$ and $p_{t}$ are the energy density, radial pressure and tangential pressure, respectively. $\zeta_{i}$ is the four-velocity and $\xi_{i}$ corresponds to a radial four-vector obeying the following relations
	\begin{equation}
		\zeta^{a}=e^{-\frac{\nu}{2}}\delta^{a}_{0}, \ \zeta^{a}\zeta_{a}=1, \ \xi^{a}=e^{-\frac{\lambda}{2}}\delta^{a}_{1}, \ \xi^{a}\xi_{a}= -1.
	\end{equation}
	
	For the metric (\ref{eq18}) and the energy-momentum  tensor (\ref{eq20}), using (\ref{eq10}), we obtain the independent set of the field equations as
	\begin{eqnarray}
		\label{eq21}
		\rho &=& -f_{Q} \left[ Q + \frac{1}{r^{2}} + \frac{e^{-\lambda(r)} \left( \lambda'(r) + \nu ' (r)\right)}{r}\right]+ \frac{f}{2},\\
		\label{eq22}    
		p_{r} &=& f_{Q} \left( Q + \frac{1}{r^{2}}\right) - \frac{f}{2},\\
		\label{eq23}    
		p_{t} &=& f_{Q} \left[\frac{Q}{2} - e^{-\lambda(r)}\left[\left(\frac{\nu'(r)}{4}+ \frac{1}{2r}\right) \left(\nu'(r) - \lambda' (r) \right) + \frac{\nu '' (r)}{2} \right] \right]- \frac{f}{2},\\
		\label{eq24}     
		0 &=& \frac{\cot \theta}{2} Q' f_{QQ}.
	\end{eqnarray}
	
	In order to close the system of equations (\ref{eq21})--(\ref{eq23}), it is necessary to prescribe a functional form of $f(Q)$. Earlier, Wang {\em et al}~\cite{wang} demonstrated that the exterior solution to the field equations corresponds to the Schwarzschild (anti-)de Sitter metric \emph{if and only if} $f_{QQ}=0$. This condition, therefore, fixes the admissible form of $f(Q)$ when constructing self-gravitating compact star models. Imposition of the condition $f_{QQ}=0$ directly leads to a linear form
	\begin{equation}
		f(Q) = \alpha Q + \beta,
		\label{eq25}
	\end{equation}
	where $\alpha$ and $\beta$ are constants. For $\alpha = -1$, we regain the GTR limit and the parameter $\beta$ is related to the cosmological constant through the relation  $\beta=2\alpha \Lambda$, in this construction.

	Note that equation (\ref{eq24}) is identically satisfied for our assumed form of $f(Q)$. Thus, in addition to the constants $\alpha$ and $\beta$, we have a system of three independent equations (\ref{eq21})-(\ref{eq23}) with five unknowns, namely $\rho(r)$, $p_r(r)$, $p_t(r)$, $\lambda(r)$ and $\nu(r)$. Hence, we are left with two degrees of freedom. To obtain a closed form solution, we first invoke the Karmarkar condition \cite{Karmakar1948}, which establishes a relationship between the metric functions and their derivatives. Earlier, many investigators have used Karmarkar condition for developing realistic stellar models \cite{Abbas2019, Sharif2022, Mustafa2023,Singh2016, Singh2016_2}. When a four dimensional curved space-time satisfies the Karmarkar condition, it can be embedded to a five dimensional pseudo-Euclidean flat space. The embedding provides an additional differential equation which subsequently permits one to generate a new class of solutions known as Class-I solutions. In this paper, we adopt the Karmarkar condition to close the system in terms of one generating function $\lambda(r)$. It should be stressed that an isotropic distribution governed by a geometry of embedding class-I in general relativity admits only three possible solutions: flat geometry, the Schwarzschild solution and the Kohler-Chao solution~\cite{singhclass1}. The above cases are highly idealized, however, by incorporating anisotropy in the matter distribution, it is possible to generate a wide class of solutions within the class-I framework. The physical motivation for the choice of anisotropy has already been discussed earlier.
	
	The Karmarkar condition can be translated to the form
	\begin{equation}
		R_{1010} \ R_{2323} = R_{1212} \ R_{3030} - R_{2102} \ R_{3103}.
		\label{eq26}
	\end{equation}
	We evaluate the non-zero Riemann tensors for our assumed line element (\ref{eq18}) as
	\begin{eqnarray*}
		R_{1010}&=& -\frac{e^{-\nu}}{4}\left[\nu^{'}\lambda^{'}-\nu^{'2}-2 \nu^{''}\right], \\
		R_{2323}&=& -r^{2} \sin^{2} \theta \left[1-e^{-\lambda}\right],\\
		R_{1212}&=& \frac{r\lambda^{'}}{2},\\
		R_{3030}&=& \frac{e^{-\nu}}{2}\left[r \sin^{2}\theta \ \nu^{'}e^{-\lambda}\right],\\
		R_{2102}&=&0,\\
		R_{3103}&=& 0.					
	\end{eqnarray*}
	Substituting these values in equation (\ref{eq26}), we obtain
	\begin{equation}
		\left[ \lambda'(r)-\nu'(r)\right] \nu'(r) \ e^{\lambda(r)} + 2 \left( 1 - e^{\lambda(r)}\right) \ \nu''(r) + \nu'^{2}(r) = 0.
		\label{eq27}
	\end{equation}
	Integration of (\ref{eq27}) yields
	\begin{equation}
		e^{\nu(r)} = \left( C + D \int \sqrt{e^{\lambda(r)}-1} \ dr \right)^{2},
		\label{eq28}
	\end{equation}
	where $C$ and $D$ are constants of integration. Thus, the line element ($\ref{eq18}$) eventually takes the form
	\begin{eqnarray}
		ds^{2} &=& -\left( C + D \int \sqrt{e^{\lambda(r)}-1} \ dr \right)^{2}  dt^{2} + e^{\lambda(r)}dr^{2} + r^{2} \left(\ d\theta^{2} + sin^{2}\theta \ d\phi^{2}\right).
		\label{eq29}
	\end{eqnarray}
	As we close the system in terms of one generating function $\lambda(r)$, we are in a position to choose a suitable form of the metric potential $\lambda (r)$ to facilitate physical analysis of the resultant configuration. In the following section, we assume a particular geometrically motivated form of the metric potential $\lambda(r)$ which has the potential of providing physically meaningful stellar model so as to examine the physical features of the resultant stellar configuration within the domain of $f(Q)$ gravity.
	
	\section{A particular model}
	\label{sec3}
	
	In the previous section, we have closed the system in terms of a single generating function $\lambda(r)$. A specific stellar model can be developed by choosing the function $\lambda(r)$ appropriately. In this paper, we invoke the Vaidya and Tikekar (VT) metric ansatz \cite{Vaidya1982} which has been found to be very useful to model realistic astrophysical objects. The VT ansatz is motivated by a geometric property that $t=$ constant hypersurface of the associated space-time, when embedded in a $4$-Euclidean space is not spherical but spheroidal. The parameter $K$, which appears in the ansatz, indicates a departure from the sphericity of the associated $3$-space. For $K=0$, the geometry is spherical. In the past, the ansatz has been utilized many investigators to model and study a wide variety of static compact stars (see e.g., \refcite{Sharma2007,Sharma2001,Sharma2002,Karmakar2007,Komathiraj2007,Kumar2014,Chattopadhyay2012,Paul2014} and references therein) and radiating stars (see e.g., \refcite{Sharma2012_2,Vaidya1996,Sarwe2010,Sharma2012}). The ansatz has also found its application in higher dimensional studies (see e.g., \refcite{Khug,Chanda2024}.)
	
	The VT ansatz is given by
	\begin{equation}
		e^{\lambda(r)} = \frac{1 + K \frac{r^{2}}{L^{2}}}{1 - \frac{r^{2}}{L^{2}}}.
		\label{eq30}
	\end{equation}
	The geometry of the associated spacetime is governed by the dimensionless curvature parameters $K$ and another parameter $L$ having the dimension of a $[length]$. Substituting (\ref{eq30}) in (\ref{eq28}) and integrating, we obtain 
	\begin{equation}
		e^{\nu(r)} = [ C - D  \sqrt{(1+K)(L^{2}-r^{2})} ]^{2},
		\label{eq31}
	\end{equation}
	where $C$ and $D$ are constants which can be determined from the boundary conditions. Subsequently, the physical variables are obtained in the form 
	\begin{eqnarray}
		\label{eq32}
		\rho &=& \frac{\beta}{2} - \frac{(K+1)(3L^{2}+Kr^{2})\alpha}{(L^{2}+Kr^{2})^{2}},\\
		\label{eq33}
		p_{r} &=& \frac{1}{2(L^{2}+Kr^{2})(D(1+K)-C\sqrt{\frac{1+K}{{L^{2}-r^{2}}}})}\left( C\sqrt{\frac{1+K}{L^{2}-r^{2}}} \right.\nonumber\\
		&&(-2\alpha (1+K) + \beta(L^{2}+Kr^{2})) - D(1+K)(-2\alpha (3+K) + \beta(L^{2}+Kr^{2})) \Biggr),\\
		\label{eq34}
		p_{t} &=& \frac{1}{2(L^{2}+Kr^{2})^{2}(D(1+K)-C\sqrt{\frac{1+K}{{L^{2}-r^{2}}}})}\left( C\sqrt{\frac{1+K}{L^{2}-r^{2}}}\right.\nonumber \\
		&& (-2\alpha L^{2}(1+K) + \beta(L^{2}+Kr^{2})^{2})
		- D(1+K)(-2\alpha (3+K) + \beta(L^{2}+Kr^{2}))\Biggr),\nonumber\\
		\label{eq35}\\
		\Delta &=& p_{t}-p_{r}.
	\end{eqnarray}
	
	The mass contained within a radius $r$ is obtained as
	\begin{equation}
		m(r) = \left(\frac{r^{3}}{12}\right)\left(\beta -\frac{6\alpha (K+1)}{L^{2}+Kr^{2}}\right),
		\label{eq36}
	\end{equation}
	which shows that the total mass increases linearly with $\alpha$.
	
	\section{Exterior space-time and boundary conditions}
	\label{sec4}
	
	The exterior space-time has the standard Schwarzschild de-Sitter form
	\begin{eqnarray}
		ds^{2} &=& -\left( 1 - \frac{2M}{r}-\frac{\Lambda}{3}r^{2}\right)dt^{2}+\left( 1 
		- \frac{2M}{r}-\frac{\Lambda}{3}r^{2}\right)^{-1}dt^{2} +r^{2}d\Omega^2, 
		\label{eq37}
	\end{eqnarray}
	where $\Lambda$ is the cosmological constant and $M$ is the total mass of a gravitating body. In $f(Q)$ gravity, the constant $\beta$ turns out to be $\beta=2\alpha \Lambda$ \cite{Mustafa2023}. Even though the cosmological constant might be relevant in a cosmological scenario, for a localized body, its effect is expected to be negligibly small ($\Lambda \sim 10^{-54} /km^{2}$). Hence, we set $\beta \approx 0$ in our construction and subsequently the exterior space-time of the stellar configuration under consideration will be described by the Schwarzschild metric
	\begin{equation}
		ds^{2} = -\left( 1 - \frac{2M}{r}\right)dt^{2}+\left( 1 
		- \frac{2M}{r}-\right)^{-1}dt^{2} +r^{2}d\Omega^2. 
	\end{equation}	
The interior solution should match the exterior Schwarzschild solution across the physical boundary $r=R$ of the star. Moreover, the radial pressure should vanish at the boundary $r=R$. The junction conditions
	\begin{eqnarray}
		\label{eq38}
		1-\frac{2M}{R}&=& e^{\nu(R)}=e^{-\lambda(R)},\\
		\label{eq39}
		p_{r}|_{r=R}&=&0,
	\end{eqnarray}
determine the constants of the model as
	\begin{eqnarray}
		C &=& \sqrt{1-\frac{2M}{R}} \left( \frac{3+K}{2}\right),\\
		D &=& \frac{1}{\sqrt{(1+K)(L^{2}+KR^{2})}}\left(\frac{1+K}{2}\right),\\
		L &=& \frac{R}{\sqrt{2M}}\sqrt{R(1+K)-2KM}.
		\label{eq40to42}
	\end{eqnarray}
	Thus, for a given mass and radius, values of the constants can be fixed for different choices of $\alpha$. This will be taken up in the following section.
	
	\section{Physical features}
	\label{sec5}
For a given value of the curvature parameter $K$, the values of the constants can be fixed if the mass and radius of the star is known. In order to analyze physical features of the resultant stellar configuration, we consider the observational data of the pulsar \textit{4U1608-52}. The mass and radius of the pulsar are estimated to be $M=1.74\pm 0.14 \ M_{\odot}$ and $R = 9.52 \pm 0.15 \ km $, respectively \cite{Gangopadhyay2013}. For the estimated mass and radius of the pulsar \textit{4U1608-52}, assuming the curvature parameter $K=50$, we determine constants  as $L=63.5691~km$, $C=17.9892$ and $D=0.0386$. For this set of values and a given value of $\alpha$, we are now in a position to analyze the physical features of the model. We note that the case $\alpha= -1$ corresponds to an anisotropic stellar configuration in standard GTR and hence, to analyze the impacts of $f(Q)$ gravity, we should consider values of $\alpha$ other than $\alpha = -1$. Note that to ensure that the physical quantities like energy density and total mass remain positive, $\alpha$ should take negative values only in this model.

\begin{figure}[ht]
			\centerline{\includegraphics[width=4.5in]{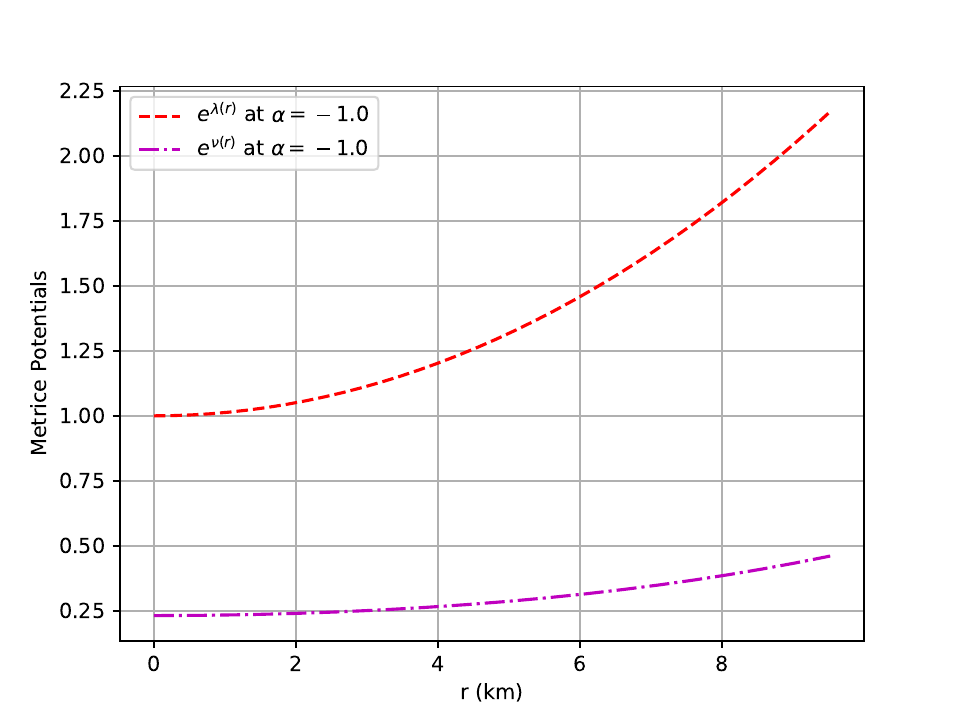}}
		\vspace*{8pt}
		\caption{ Metric potentials $e^{\lambda(r)}$ and $e^{\nu(r)}$ plotted against radial coordinate $r$. The metric potentials are regular throughout the interior of the star.}
		\label{fig1a}
		\end{figure}

\begin{figure}[ht]
			\centerline{\includegraphics[width=4.5in]{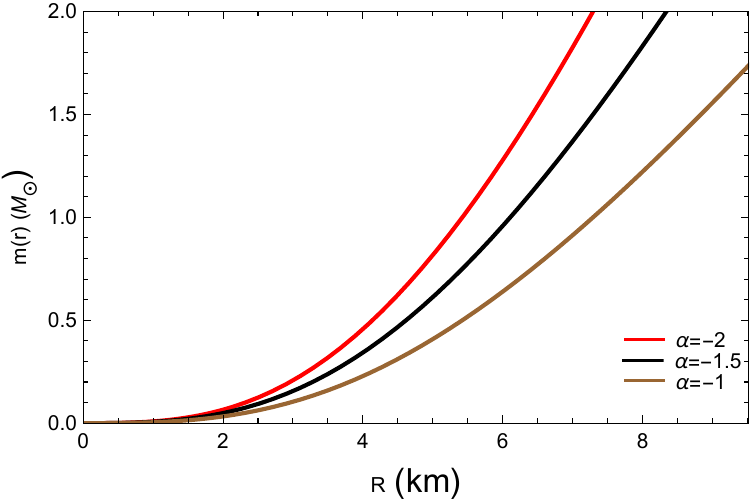}}
			\vspace*{8pt}
			\caption{Plot of mass function $m(r)$ against radial distance $r$ for different values of $\alpha$. Mass $m(r)$ within a radial distance $r$ increases in linear $f(Q)$ gravity. Note that we have used the conversion $1~M_{\odot} = 1.475~$km to express the mass function $m(r)$ in terms of $M_\odot$.}
			\label{fig1b}
		\end{figure}

In figure~\ref{fig1a}, we show the nature of the metric potentials at the interior of the star which is regular throughout the stellar configuration. Interestingly, the metric potentials remain unaltered for different choices of $\alpha$. To examine the $f(Q)$ modification on other physical quantities, we show the radial variation of physically meaningful quantities for different choices of $\alpha$. We note that the mass function $m(r)$ increases for any departure from the GTR limit ($\alpha = -1$) as shown in figure~\ref{fig1b}. This behaviour is consistent with an earlier observation made by Lin and Zhai \cite{Lin2022}.

		\begin{figure}[ht]
		\centerline{\includegraphics[width=4.5in]{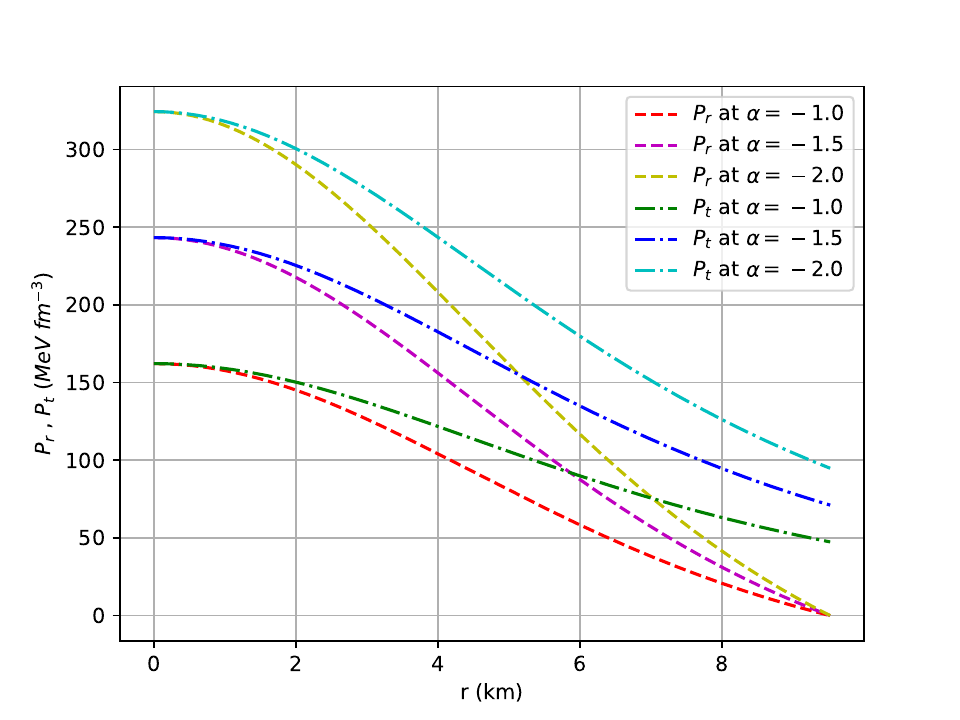}}
		\vspace*{8pt}
		\caption{Radial ($p_{r}$) and transverse ($p_{t}$) pressure plotted against radial coordinate $r$ for different values of $\alpha$. The two pressures take higher values, particularly in the core region, in $f(Q)$ gravity.}
		\label{fig2a}
	\end{figure}

		\begin{figure}[ht]
		\centerline{\includegraphics[width=4.50in]{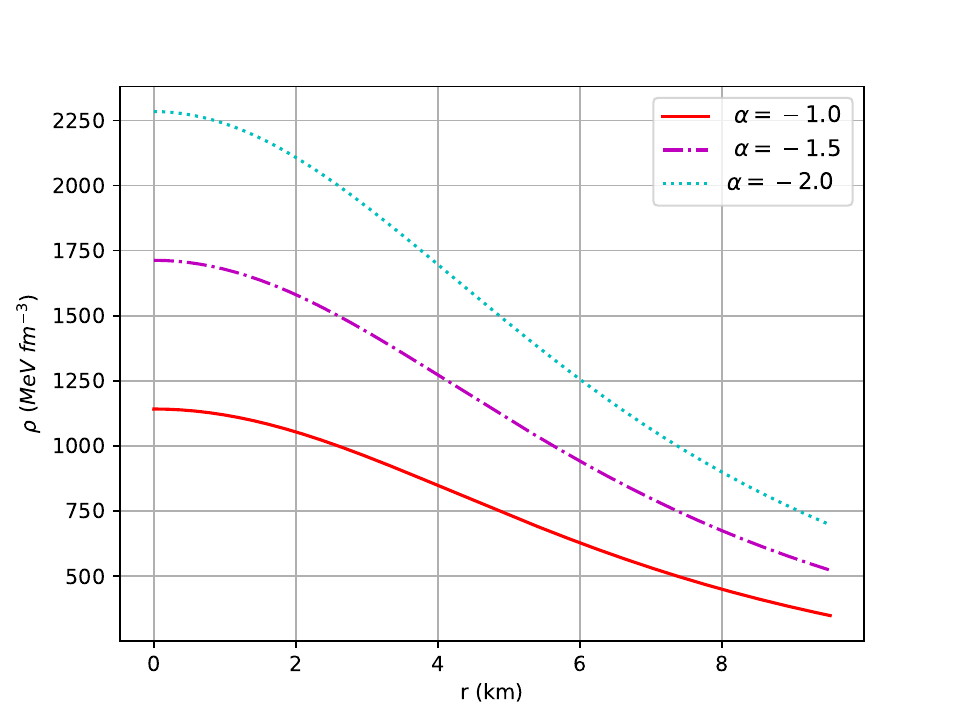}}
		\vspace*{8pt}
		\caption{Energy density($\rho$) plotted against the radial  distance $r$ for different values of $\alpha$. The energy density takes higher values in $f(Q)$ gravity.}
		\label{fig2b}
	\end{figure}

\begin{figure}[ht]
		\centerline{\includegraphics[width=4.50in]{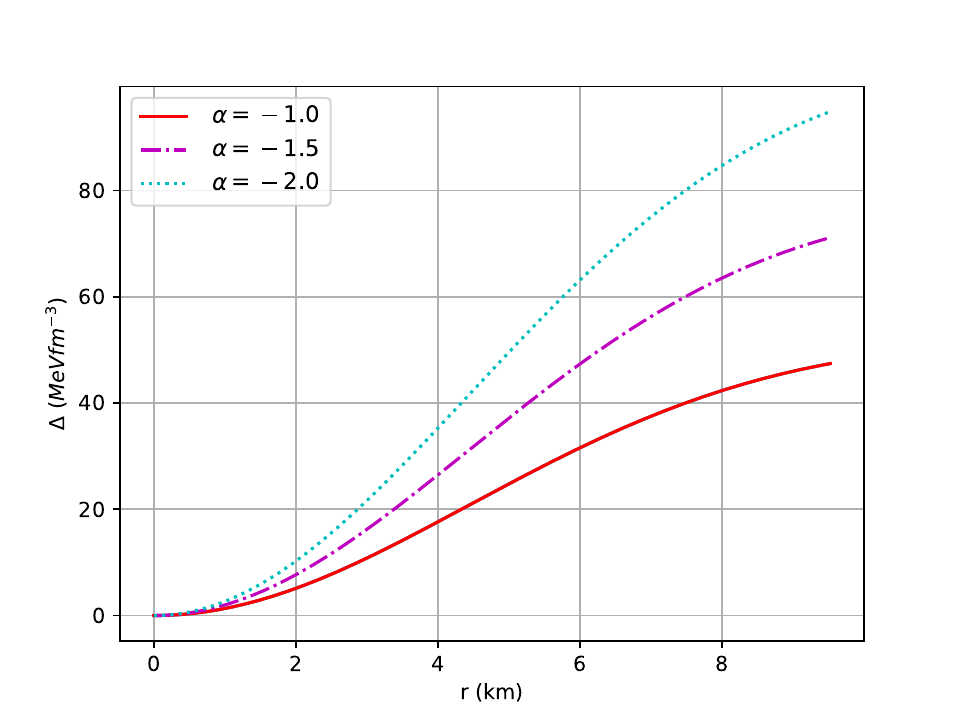}}
		\vspace*{8pt}
		\caption{Anisotropic factor  $\Delta$ plotted against radial distance $r$ for different values of $\alpha$. The anisotropy vanishes at the centre for all values of $\alpha$ and takes comparatively higher values in $f(Q)$ gravity.}
		\label{fig3a}
	\end{figure}

\begin{figure}[ht]
		\centerline{\includegraphics[width=4.50in]{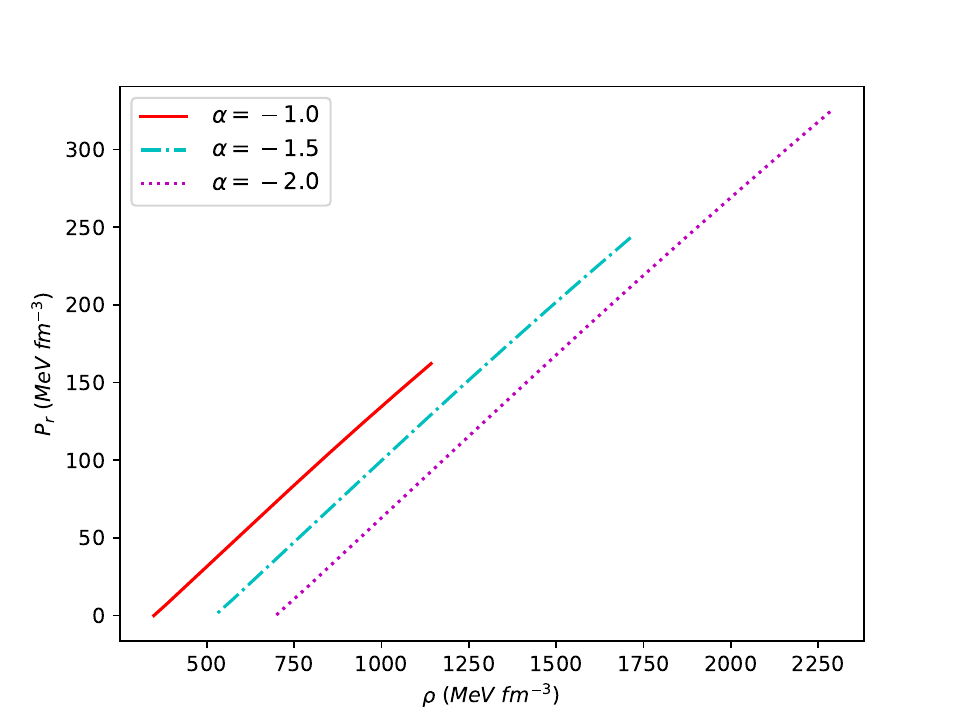}}
		\vspace*{8pt}
		\caption{Equation of state (EOS) for different values of $\alpha$. The slope of each EOS remains unaltered for different choices of $\alpha$ as the matter segment in the Einstein field equations remains the same in $f(Q)$ gravity.}
		\label{fig3b}
	\end{figure}

We note that the energy density and both radial and tangential pressures increase as $\alpha$ become more negative (figure \ref{fig2a} and figure \ref{fig2b}) i.e., in $f(Q)$ gravity the energy density and the two pressures take high values compared to standard GTR values ($\alpha = -1$). The anisotropic parameter  $\Delta$ also increases as absolute value of $\alpha$ is increased from its GR limiting value (figure~\ref{fig3a}). The thermodynamic relation between the energy density and the radial pressure provides the equation of state (EOS) of the matter composition as shown in figure~\ref{fig3b}. The EOS is almost linear and remains so for different choices of $\alpha$. It is noteworthy that for $\alpha > 0$, we get unrealistic energy density and pressure profiles and hence, the analysis is restricted to negative values of $\alpha$ only.
	
	\section{Maximum compactness bound in $f(Q)$ gravity}
	\label{sec6}
	\subsection{Compactness bound}
    Taking note of the fact that a constant density star cannot be compressed any further, in GTR, it is possible to arrive at the maximum compactness bound $M/R\leq 4/9$, the well-known Buchdahl bound \cite{Buchdahl1959}. It turns out that this bound holds for inhomogeneous stellar models as well. Impacts of anisotropy \cite{Mustafa2023,Sharma2021_3,Thirukkanesh2023, Dadhich2020,Sharma2021_4}, charge \cite{Mak2001,Bohmer2007,Bhattacharya2024} or dimension \cite{Chanda2024} on the compactness bound are well studied in GTR. In a recent article, Bhattacharya \emph{et al} \cite{Bhattacharya2024} have shown that for a charged stellar object in $f(R,T)$ gravity with $f(R,T) = R +2\chi T$, the maximum compactness bound can be obtained in the form
	\begin{equation}
		\frac{M}{R} = \frac{\frac{32}{9}\left[\frac{2}{\left(8\pi - \chi\right)}-\frac{\chi}{\left(8\pi- \chi\right)^{2}}\right]}{1+\frac{4\pi}{\left(8\pi-\chi\right)}\sqrt{\frac{16Q^2/R^2}{9}\left(\frac{3\chi}{8\pi}-2\right)+\left(\frac{\chi}{4\pi}-2\right)^{2}}},
	\end{equation}
	where $\chi$ is a dimensionless coupling parameter and $Q$ is the total charge within a sphere of radius $R$. In GTR, for a charged compact star, the above bound reduces to
	\begin{equation}
		\frac{M}{R} = \frac{8/9}{1+\sqrt{1-\frac{8Q^2/R^2}{9}}},
	\end{equation} 
which is exactly the same as the bound obtained earlier independently by Sharma \emph{et al} \cite{Sharma2021_4} and Giuliani and Rothman \cite{Giuliani2008}. Obviously, for a charge neutral star ($Q = 0$), one regains the Buchdahl bound. 
	
\begin{figure}[ht]
		\centerline{\includegraphics[width=4.50in]{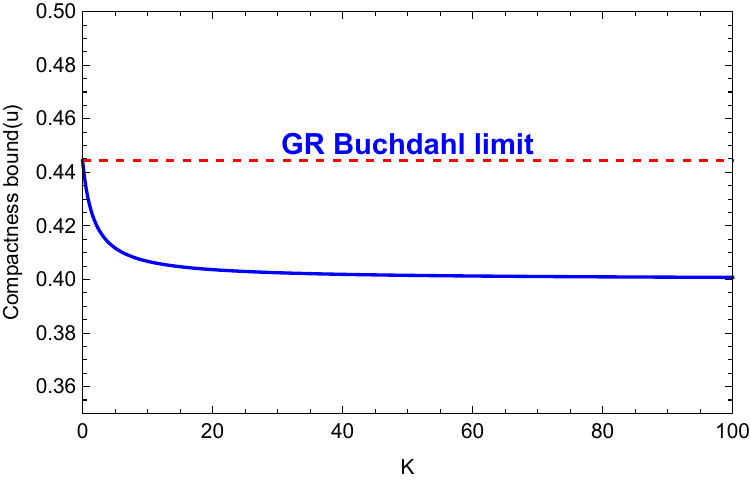}}
		\vspace*{8pt}
		\caption{Maximum compactness bound $u = M/R$ for different values of the curvature parameter $K$. The compactness bound remains below the Buchdahl bound shown by the dotted line. As $K \rightarrow 0$, one regains the Buchdahl bound. Note that in Eq~(\ref{bound}), the ratio of $M$ and $R$ is dimensionless.}
		\label{fig4}
	\end{figure}

In this paper, we intent to find the maximum compactness bound for an anisotropic stellar configuration in $f(Q)$ gravity. This can be easily obtained by noting that the central pressure must not diverse. Consequently, in equation ($\ref{eq33}$), we note that for finite central pressure we must have
	\begin{equation}
		\label{eq43}
		LD\sqrt{1+K}-C \geq 0 \ \ (L\neq 0).
	\end{equation}
	Substituting the values of $C,~D$ and $L$ in equation (\ref{eq43}), we obtain the compactness bound in $f(Q)$ gravity in the form
	\begin{equation}
		\label{bound}
	u=	\frac{M}{R} \leq \frac{2(K+2)}{5K+9},  
	\end{equation}
	 It is interesting to note that, even though the total mass increases linearly with $\alpha$ as can be seen in equation (\ref{eq36}), the maximum compactness bound is independent of $\alpha$ in our linear $f(Q)$ gravity model. In equation (\ref{bound}), substituting $K=0$, we regain the Buchdahl bound $u=\frac{M}{R} \leq \frac{4}{9}$ which is expected since in $K=0$ case, the  associated geometry becomes spherical and the model corresponds to the Schwarzschild interior solution describing an isotropic and homogeneous fluid sphere which readily provides the Buchdahl bound. It should be pointed out that for $K \neq 0$, the maximum compactness bound obtained in linear $f(Q)$ is identical to one obtained earlier by Sharma {\em et al} \cite{Sharma2021_3} for an anisotropic stellar configuration within the domain of GTR. Figure~\ref{fig4} shows variation of the compactness bound for different values of $K$. The bound never exceeds the Buchdahl bound, shown by the dotted line in the figure.
	
	\subsection{Mass-radius ($M-R$) relationship}
\begin{figure}[ht]
     	\centerline{\includegraphics[width=6.0in]{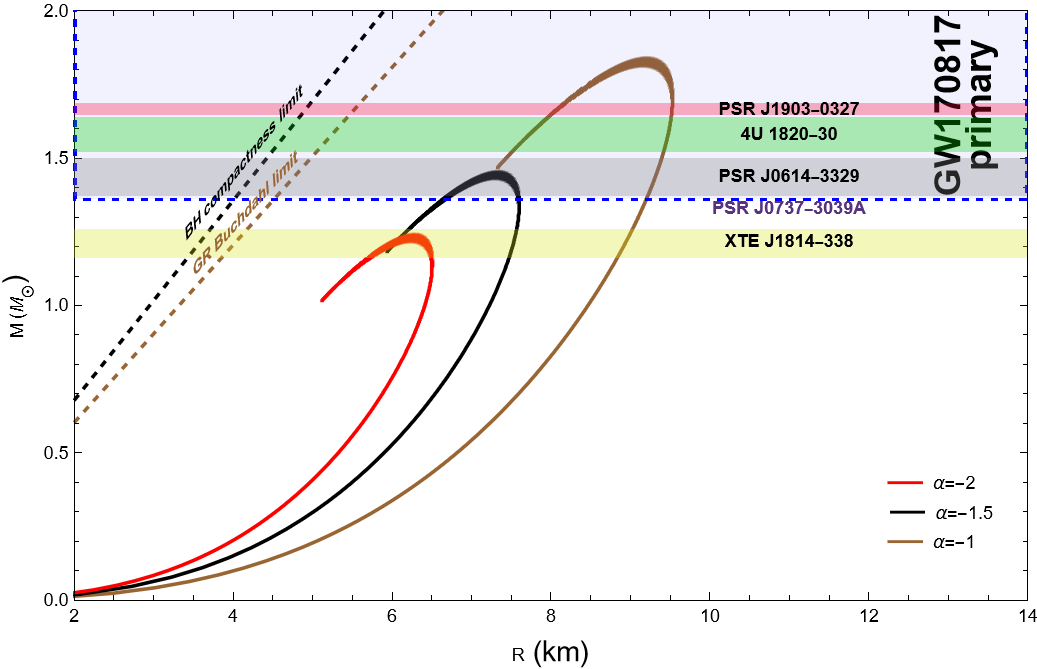}}
     	\vspace*{8pt}
     	\caption{Mass-radius ($M-R$) relationship for different values of $\alpha$ in linear $f(Q)$ gravity. As the absolute value of $|\alpha|$ is increased, the $M-R$ plot shifts towards left thereby yielding stellar configurations of smaller radii and masses. Numerical calculation shows that the compactness, however, increases as $\alpha$ becomes more negative from its GTR limit $\alpha = -1$. We have used the conversion $1~M_{\odot} = 1.475~$km to express the total mass $M$ in terms of $M_\odot$.}
     	\label{figmr}
     \end{figure}

To examine the impact of $f(Q)$ modification in the $M-R$ relationship, we integrate the  Tolman-Oppenheimer-Volkoff (TOV) equations
    \begin{eqnarray}
    \frac{dp_r}{dr}
    &=& -(\rho + p_r)\,
    \frac{m + 4\pi r^{3} p_r}{r(r - 2m)}
    + \frac{2}{r}(p_t - p_r),\label{tov_aniso}\\
    m(r) &=& 4\pi \int_{0}^{r} \rho(\tilde r)\, \tilde r^{2}\, d\tilde r, \label{mass}
    \end{eqnarray} 
for a given curvature parameter $K$ by considering different values of $\alpha$. Since the EOS is not furnished a priori in our construction, we utilize equations (\ref{eq32}) -(\ref{eq35}) to numerically integrate the TOV equations. The $M-R$ relationship for different values of $\alpha$ is shown in figure~\ref{figmr}. For numerical calculation, we assume $K=2$ and consider three distinct values of $\alpha$ including the GTR limit $\alpha = -1$.\\
In the figure, we note that for $\alpha=$ $-1.0, -1.5, - 2.0$, the respective values of the maximum masses are obtained as  $1.8277\ M_{\odot},~1.4438\ M_{\odot}$ and $ 1.2305\ M_{\odot}$, respectively. It is important to note that even though an increasing value of  $\left|\alpha\right|$  raises the central density and pressures relative to GR, the dominant gravitational term in the TOV equations, $(\rho + p_{r})\left(m + 4\pi r^{3} p_{r}\right)$ scales as $\propto \alpha^{2}$ and the pressure gradient support scales only as $\propto \left|\alpha\right|$. Consequently, self gravity grows faster than pressure support with increasing $\left|\alpha\right|$. This leads to a reduction in the maximum stable mass and radius despite the larger local densities.
     
Our analysis shows that the  stellar model developed in this paper by assuming a linear modification in the function $f(Q)$ is more suitable for the description of lighter ($M < 2M_\odot$) and ultra-compact stars such as $XTE~J1814-338$, with an estimated mass $M=1.21\pm0.05$ and radius $R=7.0\pm0.4~km$~\cite{kini}.The estimated radii for several such selected compact objects are presented in Table~\ref{Table3}, while the corresponding observed mass bands are shown in figure~\ref{figmr}.
\begin{table}[ht]
\tbl{Estimated radii of selected compact objects for different values of $\alpha$.\label{Table3}}{
\tabcolsep13pt
\begin{tabular}{@{}ccccc@{}}
\hline
Compact object & $M_\odot$ & $R_{\text{observed}}$ (km) & $\alpha$ & $R_{\text{pred}}$ (km) \\ 
\hline
               &            &            & $-1.0$ & $8.93$ \\ 
XTE~$J1814-338$~\cite{kini} & $1.21\pm0.05$ & $7.0\pm0.4$ & $-1.5$ & $7.53$ \\ 
               &            &            & $-2.0$ & $6.43$ \\
\hline
               &            &            & $-1.0$ & 9.31 \\ 
PSR~$J0614-3329$~\cite{Mauviard2025} & $1.44^{+0.06}_{-0.07}$ & $10.29^{+1.01}_{-0.86}$ & $-1.5$ &  7.31\\ 
               &            &            & $-2.0$ & -- \\
\hline
               &            &            & $-1.0$ & 9.52 \\ 
PSR~$J1903+0327$~\cite{Freire2011} & $1.667\pm0.021$ & --- & $-1.5$ & --- \\ 
               &            &            & $-2.0$ & --- \\
\hline
               &            &            & $-1.0$ & 9.14 \\ 
PSR~$J0737-3039A$~\cite{Kramer2006} & $1.3381\pm0.0007$ & -- & $-1.5$ & 7.59 \\ 
               &            &            & $-2.0$ & --- \\
\hline
               &            &            & $-1.0$ & 9.46 \\ 
$4U~1820-30$~\cite{Guver2010} & $1.58\pm0.06$ & $9.11\pm0.40$ & $-1.5$ & --- \\ 
               &            &            & $-2.0$ & --- \\
\hline
               &            &            & $-1.0$ & $\ge 9.19$ \\ 
GW170817 (primary)~\cite{Abbott2017} & $1.36$--$2.26$ & --- & $-1.5$ & $\ge 7.58$ \\ 
               &            &            & $-2.0$ & --- \\
\hline

\end{tabular}}
\end{table}

In figure~\ref{figmr} and Table~\ref{Table3}, we note that most of the stars can well be described in the GTR limit with $(\alpha=-1)$, where the theoretically predicted radii of pulsars such as PSR~$J0614-3329$~\cite{Mauviard2025} and $4U~1820--30$~\cite{Guver2010} are aligned with  with observational results. However, for pulsars like  XTE~$J1814-338$, the theoretically predicted radius in the pure GR limit is noticeably larger than its observed estimate. Interestingly, in our $f(Q)$ gravity inspired model with $\alpha=-1.5$, the predicted radius is in better agreement with the observed results. In other words, the parameter $\alpha$ can be used to fine-tune the stellar observables.  It should, however, be stressed that as the magnitude of $|\alpha|$ increases, the stellar compactness increases in our model. Note that while using the data from the event GW$170817$~\cite{Abbott2017}, the lower bound of the primary neutron star radius is predicted for $\alpha=-1,-1.5$ only, as for $\alpha=-2$ the total mass falls below the estimated range.

\section{Concluding Remarks}\label{sec7}
In this paper, we have constructed an anisotropic stellar model in the framework of $f(Q)$ gravity by employing the Karmarkar's condition together with the Vaidya-Tikekar ansatz for the metric potential $g_{rr}$. The adopted linear modification, $f(Q)=\alpha Q+\beta$ adequately simplifies the field equations, thereby allowing us to analyze the effects of non-metricity on the internal structure and gross physical properties of a relativistic star. By matching the interior solution to the exterior Schwarzschild metric, we have fixed values of the model parameters for a given mass and radius. Subsequently, for a range of values of $\alpha$, we have analyzed the energy density and pressure profiles which exhibit distinctive features. A notable outcome of our construction is that for more negative values of $\alpha$, the density, radial pressure, tangential pressure and the anisotropy increase in magnitude. We also note that the even though the total mass increases linearly with $\alpha$, the maximum compactness bound remains independent of $\alpha$ in this construction and never exceeds the Buchdahl bound. The maximum compactness bound is entirely governed by the Vaidya-Tikekar curvature parameter $K$. One regains the Buchdahl bound for $K=0$. This shows that the maximum permissible compactness is entirely governed by the underlying geometry encoded through $K$. In the $M-R$ plot, we note that the stable branch shifts toward more compact and comparatively lower masses and radii as the magnitude of $\alpha$ increases. As the linear $f(Q)$ modification tends to reduce the maximum mass, the model can accommodate ultra-compact objects. We have illustrated how the observed radius of the pulsar $XTE~J1814-338$ can be fine tuned by the parameter $\alpha$ in our linear $f(Q)$ gravity stellar model.

A comparative look at relativistic stellar models reveals that, despite differences in micro-physics and gravitational extensions, the upper bound on compactness essentially lies below the Buchdahl bound across all the frameworks. In $f(R,T)$ gravity, both the embedding class-I analysis of strange stars \cite{MauryaPRD2019} and interacting quark-matter configurations \cite{JHEAp2024}, the total mass increases for a fixed central density, however, the maximum compactness remains below the Buchdahl limit. Non-conservative uni-modular and trace-free geometries generate heavier polytropes without violating the compactness bound \cite{DarkUnimod2024}. Similarly,  Brans-Dicke models with massive scalar fields \cite{BD2022} and standard GTR extensions based on linear, dark energy or Chaplygin equations of state \cite{EPJC2019Strange,EPJC2020,FoundPhys2019,MPLA2019Shell,EPJP2017,MPLA2019Chaplygin} also provide $M/R <4/9$. Embedding-based analyses in $f(R,T)$ gravity \cite{CPC2020} and GR \cite{EPJC2021} predict massive objects within the Buchdahl bound, as do generalized van der Waals \cite{EPJC2022VDW}, $f(R,T)$ linear-coupling models \cite{CJPH2021} and charged quark-matter stars in GTR \cite{PhysScr2022}. In rainbow gravity with color-flavor-locked quark matter, although very massive configurations may be realized, the resulting models nevertheless satisfy the Buchdahl constraint \cite{PDU2025}. It is interesting to note that in the charged stellar model developed within the framework of $f(T)$ gravity by  Singh \textit{et al},  the compactness bound might exceed the Buchdahl bound~\cite{singhft}. A similar observation is also noticed in the work of Maurya {\em et al} \cite{Maurya_Forts} where a charged compact stellar model has been developed in $f(Q)$ gravity. However, in our model, the maximum compactness bound remains unaltered by the $f(Q)$ modification in the linear regime. We note a shift in the mass-radius curve towards more compact and comparatively smaller mass stellar configurations, however, the maximum compactness remains strictly below the Buchdahl limit.

It should be stressed that the current investigation is restricted to the linear form of $f(Q)$ where the solution has been generated via the Karmarkar's embedding approach. Moreover, in the assumed form $f(Q) = \alpha Q+ \beta$, as the cosmological constant is positive but small, we effectively have $\beta \approx 0$. In a recent article \cite{Jensko2024}, it has been argued  that if the assumed coordinate system is not compatible with the coincident gauge, the choice of $f(Q) = \alpha Q+ \beta$ would reduce the theory to the standard GTR results in the presence of of a cosmological constant. This is reflected in our results as well. Obviously, one needs to study non-trivial models in $f(Q)$ gravity. A different form of  the function $f(Q)$ may reveal a broader range of stellar properties. Such extensions should clarify the impacts of non-metricity on stellar properties and maximum permissible mass, in particular, in ways not permitted by the linear model assumed in our analysis. Wang et al~\cite{wang} demonstrated that for a static, spherically symmetric configuration, written in the coincident gauge with a diagonal metric, the field equation forces either a linear $f(Q)$, which is dynamically equivalent to GTR with a cosmological constant or a constant $Q$ branch in vacuum. This restriction, however, applies only to the specific coordinate choice and to the exterior region where $Q$ becomes constant. Within a stellar interior, where $Q(r)$ varies non-trivially, nonlinear models with $f_{QQ}\neq 0$ can still introduce genuinely new dynamics by modifying the effective gravitational coupling and the matter-geometry interaction. Exploring such nonlinear extensions, together with alternative treatments for solution generating techniques would help us understand how robust the conclusions of the current work are. It will also be interesting to examine whether additional degrees of freedom incorporated through other factors such as rotation, heat flux etc. studied within the framework of $f(Q)$ gravity, can provide new physics in regimes inaccessible to the linear theory. All these, however, are beyond the scope of the current investigation and will be taken up elsewhere.

\section*{Acknowledgments}
RS gratefully acknowledges support from the Inter-University Centre for Astronomy and Astrophysics (IUCAA), Pune, India,  under its Visiting Research Associateship Programme.

	\section*{ORCID}
	\noindent Ranjan Sharma - \url{https://orcid.org/0000-0002-1468-7160}

\end{document}